# Ripple-assisted adsorption of noble gases on graphene at room temperature

Weilin Liu[1,†], Xianlei Huang[1,†], Li-Guo Dou[1,†], Qianglong Fang[2,†], Ang Li[3,†], Guowen Yuan[1,*], Yongjie Xu[1], Zhenjia Zhou[1], Jun Li[1], Yu Jiang[1], Zichong Huang[1], Zihao Fu[1], Peng-Xiang Hou[4], Chang Liu[4], Jinlan Wang[2,5], Wu Zhou[3,*], Ming-Gang Ju[2,*], Shao-Chun Li[1,6,*], Hui-Ming Cheng[4,7] and Libo Gao[1,8,*]

[1] National Laboratory of Solid State Microstructures, Jiangsu Key Laboratory for Nanotechnology, Jiangsu Physical Science Research Center, School of Physics, Nanjing University, Nanjing, China.

[2] Key Laboratory of Quantum Materials and Devices of Ministry of Education, School of Physics, Southeast University, Nanjing, China.

[3] School of Physical Sciences, University of Chinese Academy of Sciences, Beijing, China.

[4] Shenyang National Laboratory for Materials Sciences, Institute of Metal Research, Chinese Academy of Sciences, Shenyang, China.

[5] Suzhou Laboratory, Suzhou, China.

[6] Hefei National Laboratory, Hefei, China.

[7] Institute of Technology for Carbon Neutrality, Shenzhen Institute of Advanced Technology, Chinese Academy of Sciences, Shenzhen, China.

[8] State Key Laboratory of Chemo/Biosensing and Chemometrics, Key Laboratory for Micro-Nano Physics and Technology of Hunan Province, College of Materials Science and Engineering, Hunan University, Changsha, China.

* e-mail: gwyuan@nju.edu.cn; wuzhou@ucas.ac.cn; juming@seu.edu.cn; scli@nju.edu.cn; lbgao@nju.edu.cn

[†] Equally contributed to this work.

## ABSTRACT

Controllable gas adsorption is critical for both scientific and industrial fields, and high-capacity adsorption of gases on solid surfaces provides a significant promise due to its high-safety and low-energy consumption. However, the adsorption of nonpolar gases, particularly noble gases, poses a considerable challenge under atmospheric pressure and room temperature (RT). Here, we theoretically simulate and experimentally realize the stable adsorption of noble gases like xenon (Xe), krypton (Kr), argon (Ar), and helium (He) on highly rippled graphene at RT. The elemental characteristics of adsorbed Xe are confirmed by electron energy loss spectroscopy and X-ray photoelectron spectroscopy. The adsorbed gas atoms are crystalized with periodic arrangements. These adsorbed noble gases on graphene exhibit high stability at RT and can be completely desorbed at approximately 350 °C without damaging the intrinsic lattice of graphene. The structural and physical properties of graphene are significantly influenced by the adsorbed gas, and they fully recover after desorption. Additionally, this controllable adsorption could be generalized to other layered adsorbents such as $NbSe_2$, $MoS_2$ and carbon nanotubes. We anticipate that this ripple-assisted adsorption will not only re-define the theoretical framework of gas adsorption, but also accelerate advancements in gas storage and separation technologies, as well as enhance the applications in catalysis, surface modification, and other related fields.

**Keywords:** noble gases, stable adsorption, ripple, graphene, room temperature

## INTRODUCTION

Adsorption theory has evolved over a century from Freundlich's empirical formula to Langmuir's isothermal model and the Brunauer-Emmett-Teller (BET) theory [1-4], maturing into a well-established framework where the adsorption of gas molecules on solid surfaces is categorized into chemisorption and physisorption. Chemisorption involves high interaction energy, and the resultant chemical bonds usually lead to the degradation of adsorbents [5-8]. In contrast, physisorption is considered to be reversible with low regeneration energy consumption, the structural and chemical properties of the adsorbent are basically unchanged, and most gas molecules interact with the solid surface through the weak van der Waals (vdW) interactions. Solid-gas interface adsorption underlies numerous significant phenomena and applications, including gas storage and separation [9-12], surface catalysis [6,13,14], modification of the adsorbents, and vapour growth of crystals [15-17].

Among all gas molecules, nonpolar gases, especially noble gases such as xenon (Xe), krypton (Kr), argon (Ar), and helium (He), are challenging for stable adsorption on solid surfaces under normal conditions, *i.e.* atmospheric pressure (AP) and room temperature (RT). This difficulty arises from their fully occupied orbitals, which result in the saturated electronic charges and weak interactions with solid surfaces. Some experimental efforts have been made to improve the adsorption ability of noble gases, mainly including lowing temperature and utilizing materials with high specific surface area as adsorbents [9,10,16-23]. For instance, He atoms have been successfully adsorbed onto graphite at temperature of 120 mK [21], and Ar, Kr, and Xe atoms have also been adsorbed on surface-confined metal-organic framework (MOF) at temperatures lower than 8 K [24]. In these cases, the low temperature adsorption of noble gases is classified as physisorption. However, these adsorbed noble gas atoms are restricted to cryogenic temperatures, and the desorption occurs spontaneously once the temperature exceeds the critical thresholds. To date, stable adsorption of noble gases atoms under normal conditions has not been experimentally achieved, and theoretical guidance remains inadequate.

## RESULTS AND DISCUSSIN

### Realization of noble gas atoms adsorbing on rippled graphene

Figure 1a illustrates the schematic of adsorbing noble gas atoms on graphene, using Ar as an example. Due to the weak vdW interactions, noble Ar atoms typically cannot adsorb on graphene with absolutely flat surface under normal conditions (Scene I. Non-adsorption). However, graphene inherently exhibits intrinsic ripples [25,26], a consequence of thermodynamic stability in two-dimensional (2D) structures, which induce slight out-of-plane deformations in the $sp^2$ planar lattice, with typical amplitudes of 0.7 Å and corrugation diameter of approximately 80 Å [25]. Recent studies have suggested that these intrinsic ripples can reduce the catalytic barrier of $H_2$ molecules and dissociate them into atoms [27-29]. Considering the gas adsorption is a prerequisite for the surface catalysis, we suppose the enhanced ripples can also act to reduce the adsorption energy. Based on these insights, we demonstrate the stable adsorption of noble gas atoms on highly rippled graphene (Scene II. Adsorption), where no covalent bonds are formed between carbon and Ar atoms. Besides, due to the strong ripple perturbation at elevating temperature, the adsorbed Ar atoms will be readily desorbed from the rippled graphene (Scene III. Desorption).

Then, we theoretically calculated the total adsorption energies ($E_{ads}$) of graphene with varying ripple curvature and one adsorbed noble gas atom. The negative $E_{ads}$ values confirm the exothermic adsorption, with more negative values indicating greater energy release and higher stability. For clarity, the absolute values of $E_{ads}$ are employed to represent adsorption strength. For flat graphene system with one Ar atom, $E_{ads}$ is only 100 meV (Fig. 1b), such a value will be compensated by thermal fluctuations and interfacial energy, resulting in desorption. As the ripple curvature increases, $E_{ads}$ progressively increases. When the curvature reaches 0.116, $E_{ads}$ rises to 171 meV, indicating enhanced adsorption. This trend holds with more adsorbed atoms (Fig. S1a), confirming curvature-dependent behaviour. Furthermore, $E_{ads}$ increases as increasing the molecular weight of the adsorbed gas, *i.e.*, the stable adsorption of Xe atoms on rippled graphene occurs more readily than that of He.

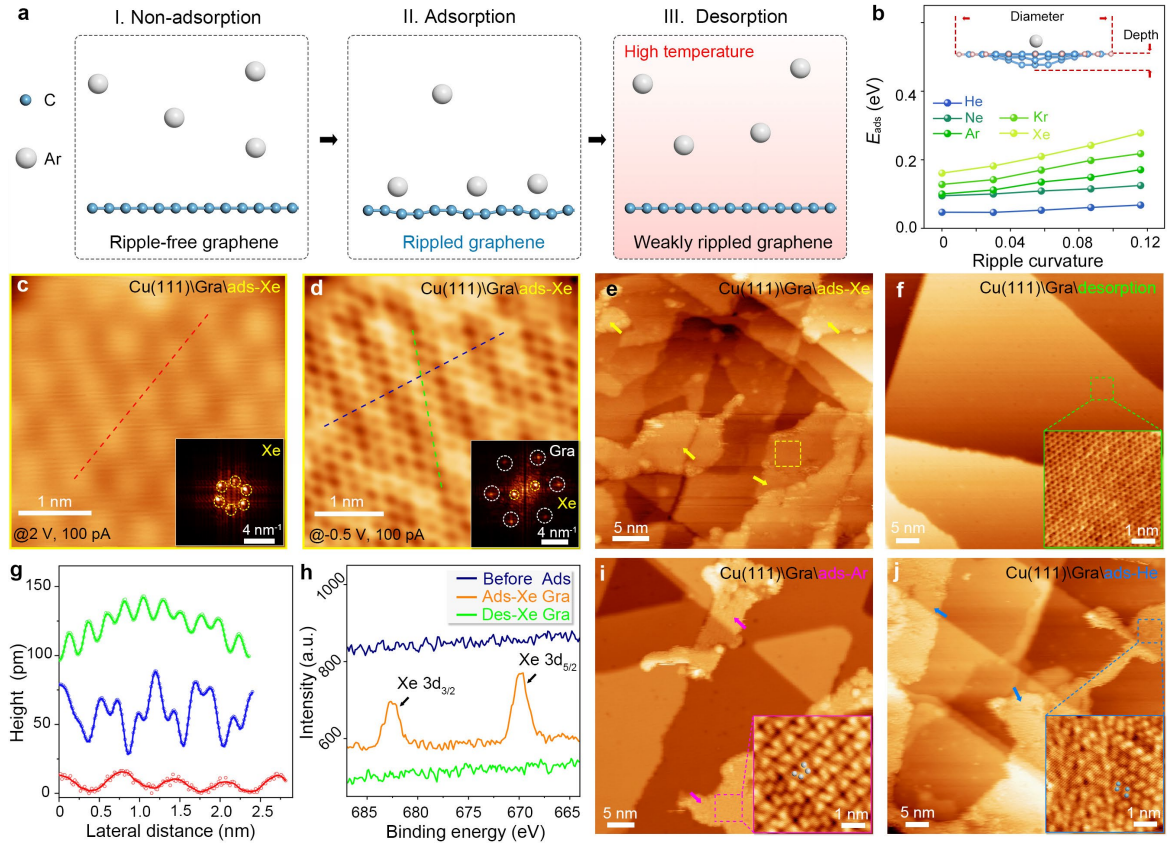

**Figure 1. Noble gas atoms adsorbed on rippled graphene.** (a) Schematic of noble gas atoms adsorbing on rippled graphene. Left (I. Non-adsorption) illustrates no Ar atoms adsorbed on absolutely flat graphene under normal conditions; Middle (II. Adsorption) illustrates the stable adsorption of Ar atoms on the highly rippled graphene; Right (III. Desorption) illustrates the desorbing Ar from graphene at high temperature. Here Ar represents the noble gas atoms. (b) Theoretical calculated $E_{ads}$ of graphene with different ripple curvature adsorbing one noble gas atom. Inset illustrates the side view of rippled graphene (96 carbon atoms) adsorbing one Ar atom. The ripple curvature is the ratio of height to lateral diameter. (c) Large-scale STM topographic image of partially ads-Xe on Cu(111)\Gra. The irregular domains labelled by arrows are formed by ads-Xe. (d, e) Typical STM topographic images of ads-Xe on Cu(111)\Gra measured under different bias voltages. They are collected from the dashed box labelled in (c). Insets are their corresponding FFT patterns. (f), Typical STM image of Cu(111)\Gra after fully desorption via thermal annealing, revealing a clean surface free of adsorbed gases. The inset shows a zoom-in image of graphene lattice, highlighting its defect-free structure. (g), Height profiles of ads-Xe graphene collected from the dashed lines in (d) and (f). The lateral distance between the neighbouring Xe atoms is ~6.8 Å. (h), Fine XPS spectra of as-transferred graphene, ads-Xe and des-Xe graphene. The Xe characteristic peaks of $3d_{3/2}$ and $3d_{5/2}$ core levels emerge. (i, j), Typical STM topographic images of partially ads-Ar (i) and ads-He (j) on Cu(111)\Gra. The irregular domains are formed by the adsorbed Ar and He atoms, and the insets are the zoom-in images of the Ar and He domains.

We experimentally utilize three different approaches to adsorb noble gases at RT. The adsorption process involves a weakly ionized plasma comprising electrons along with polarized and ionized noble gas species. Alternative methods can also yield equivalent adsorption results. Critically, those technique ensures complete preservation of the graphene lattice integrity (details in Methods section). The kinetic energy of the activated gas atoms may intensify the ripple depths of graphene, thereby achieving stable adsorption. Firstly, the scanning tunnelling microscope (STM) images of Cu(111)\monolayer (1L) graphene (Gra) film with adsorbed Xe atoms (ads-Xe) are performed under ultra-high vacuum. Figure 1c shows a typical large-scale STM image of ads-Xe on Cu(111)\Gra after a short-time 90 °C annealing. Unlike the well-defined Cu atomic steps, aggregates of ads-Xe forms irregular domains with different thickness. While zooming in an ads-Xe domain, an obviously closest packed structure of crystalized Xe is observed under the bias voltage of 2 V (Fig. 1d). The lateral distance between neighbouring Xe atoms is ~6.8 Å (Fig. 1g, red line). As decreasing the bias voltage, the lattice arrangement of Xe undergoes different deformations, while the underlying graphene honeycomb lattice becomes increasingly discernible (Fig. 1e, and Fig. S1b-f). Additionally, the ads-Xe atoms seem to keep the strict alignment with the underlying lattice orientation of graphene. The structural symmetry of ads-Xe emerges from six-fold into two-fold. The lattice spacing of graphene is measured to keep ~2.4 Å, but the aligned distance for ads-Xe is reduced to ~5.9 Å (Fig. 1g, green and blue lines). The ads-Xe can be completely removed via 350 °C annealing, leaving the defect-free hexagonal lattices of graphene (Fig. 1f), indicating the non-destructive adsorption-desorption processes.

The elemental identification of ads-Xe is further performed with X-ray photoelectron spectroscopy (XPS), and the distinct characteristic peaks of Xe $3d_{3/2}$ and $3d_{5/2}$ core levels at 682.5 and 669.8 eV appear, respectively (Fig. 1h and Fig. S2). After subsequent desorption, the Xe-related spectral features vanish completely.

Figure 1i-j display the STM images of Cu(111)\Gra with partially adsorbed Ar atoms (ads-Ar) and partially adsorbed He atoms (ads-He), respectively. They are obtained following short-time 180 °C annealing. It is worth noting that the ads-Ar and ads-He inside the irregular domains usually present a dimer configuration with the periodical array, and the lateral distances of neighbouring Ar and He atoms within a dimer are ~300 pm and ~230 pm, respectively.

Moreover, scanning tunnelling spectroscopy (STS) of ads-Xe and ads-Ar graphene are further collected (Fig. S3). The domain regions of ads-Xe and ads-Ar exhibit the distinct shoulder peaks, which is similar to the STS shoulder peak of Xe or Kr adsorbed on Au(111) at extremely low temperature [30]. The peak positions systematically shift with the annealing temperature, and there is also a small but clear bandgap of ~80 meV emerged in ads-Xe graphene.

## Noble gases adsorbed on suspending graphene

Figure 2a-d show the selected area electron diffraction (SAED) of suspending graphene (sus-Gra) films with ads-Xe at different collection stages. There is an additional set of hexagonal symmetrical SAED patterns arisen besides graphene, and these extra patterns are attributed to the crystallized ads-Xe. For these measurements, we utilize a very low dose rate of ~0.1 pA/μm$^2$ and a 12-second collection duration, but these patterns are still gradually weakened with each measurement interval. After only 48 s exposure, the extra SAED pattern of ads-Xe crystals completely disappears, indicating the adsorbed Xe atoms are easily desorbed by the 80 kV electron beam. Following complete desorption, a perfect hexagonal lattice of graphene without any crystalline defects is obtained (Fig. 2e), indicating the non-destructive feature of the adsorption-desorption process. Notably, the observed SAED pattern for adsorbed Xe atoms on graphene at RT corresponds to the lattice constant of 5.3 ± 0.2 Å, which is smaller than the value obtained by STM measured at 77 K. Additional SAED patterns collected at random locations further show that the lattice orientations of ads-Xe always align with the underlying graphene (Fig. S4).

To further confirm the existence of ads-Xe on sus-Gra, the electron energy loss spectroscopy (EELS) equipped in the scanning transmission electron microscope (STEM) is performed under the mildest conditions with 60 kV and the lowest dose rate of 0.025 pA/μm$^2$. After cumulative collection over 3600 s, a discernible Xe characteristic peak at ~699 eV is detected (Fig. 2f). However, even under the mildest adjustable conditions, we are still unable to detect the characteristic peaks from ads-Kr, ads-Ar or ads-He, probably due to their easier desorption compared with ads-Xe.

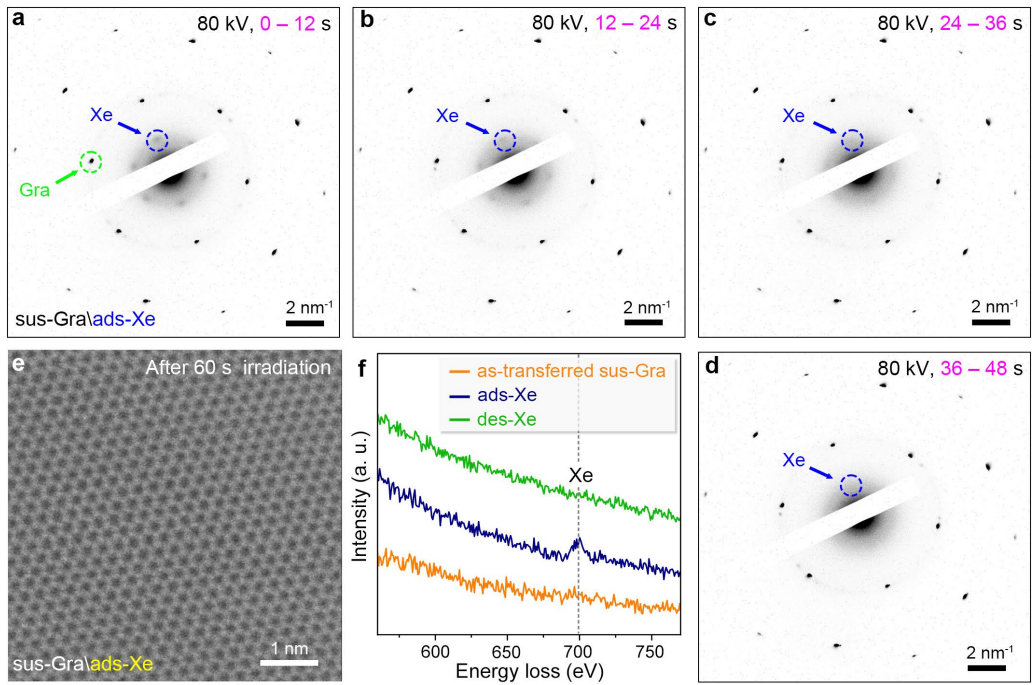

**Figure 2. STEM observations of ads-Xe on suspending graphene.** (a, d), Typical SAED patterns of sus-Gra with ads-Xe, the labelled green and blue dash circles originate from graphene and ads-Xe, respectively. The SAED patterns in (a, d) are successively captured for 12 s, and SAED patterns from ads-Xe disappear after ~36 s e-beam irradiation. The sixfold symmetrical ads-Xe patterns correspond to the in-plane lattice constant of 5.3 ± 0.2 Å. (e), Typical STEM image of sus-Gra with ads-Xe after 60 s e-beam irradiation, there are only perfect graphene lattice left. (f), EELS spectra of as-transferred sus-Gra, ads-Xe and des-Xe graphene. A distinct Xe characteristic peak located at ~699 eV is clearly observed in the ads-Xe graphene.

## Stability of the ripple-assisted adsorption

The stability of adsorbed noble gas atomic crystals is monitored by the property changes of graphene. Figure 3a plots typical Raman spectra of the as-grown graphene film on Cu(111), ads-Ar and des-Ar graphene, while Raman is highly sensitive to graphene lattice deformation and the intensity ratio of D to G peak ($I_D/I_G$) can be utilized as an indicator of non-$sp^2$ hybridization [31-34]. The as-grown graphene is defect-free, and an obvious D peak (~1350 cm$^{-1}$) appears for ads-Ar. The D peak can be fully eliminated for des-Ar by thermal annealing. This adsorption-induced D peak as well as its removal, is also applicable to other adsorbed noble gases (Fig. S5a). Given the chemical inertness of Ar atoms, the appearance of D peaks can be more plausibly attributed to the out-of-plane displacement of carbon atoms rather than the formation of C-Ar bonds, although the latter possibility cannot be ruled out, and the disappearance of the D peak corresponds to the complete desorption, which behaves different from the origins of edges, atomic vacancies and C-H bonds [33,35]. Along with the adsorption, both the G and 2D peaks exhibit a minor redshift, primarily due to the enhanced tensile strain (Fig. S5b-d).

Figure 3b plots the Raman spectra of transferred graphene (1−3 layers) on Si\SiO$_2$ after adsorbing Ar at RT. The bilayer (2L) and triple-layer (3L) graphene consist of stacked $^{12}$C and isotopically labelled $^{13}$C graphene films, named as $^{13}$C\$^{12}$C and $^{13}$C\$^{12}$C\$^{12}$C, respectively. Under the same adsorption condition, 1L $^{12}$C graphene exhibits the highest ripple deformation, with an $I_D/I_G$ ratio of ~0.61. The $I_D/I_G$ ratio decreases as increasing the layer number, and the $^{12}I_D/^{12}I_G$ ratio decrease to ~0.22 for 2L graphene, while the $^{13}I_D/^{13}I_G$ ratio is ~0.09 for 3L graphene. Noting that, the $I_D/I_G$ of $^{12}$C and $^{13}$C in different layer order both shows reduction, indicating that rippled deformations are formed in all the layers and progressively diminish as increasing layer depth. These ripple-induced D peak can also be completely recovered by thermal annealing (Fig. S5e).

Additional studies on exfoliated few-layer graphene (1–4 layers) further confirm that the $I_D/I_G$ ratio systematically weakens as increasing the layer number (Fig. S5f-g). Moreover, the adsorption process is also influenced by the substrate on which graphene is contacted. Under identical adsorption conditions, graphene on a substrate demonstrates a higher $I_D/I_G$ ratio compared to suspending graphene (Fig. S5h-j). Longer adsorption duration cannot enhance the ripple deformation once reaching its maximum, and they can be fully recovered after desorption. Furthermore, the adsorbed gas atoms on graphene are uniformly distributed across the whole wafer and remain highly stable, maintaining nearly unchanged even under vacuum for 5-months at RT (Fig. 3c and Fig. S6).

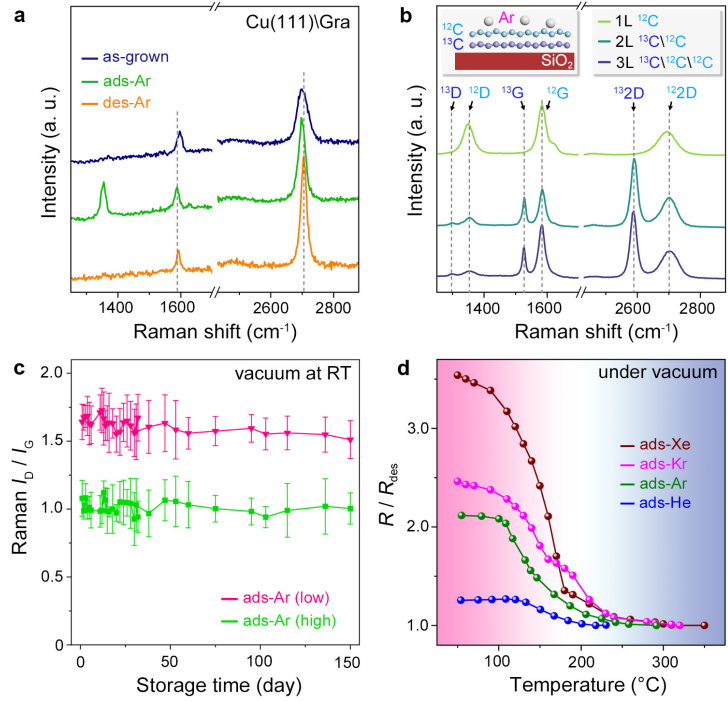

**Figure 3. Structural changes of graphene upon noble gases adsorption.** (a), Raman spectra of as-grown, ads-Ar and des-Ar graphene on Cu(111). The ads-Ar induce the $sp^3$ phase transition (D peak) in graphene and the rippled deformation can be completely recovered. (b), Raman spectra of graphene with different layer number adsorbing Ar. The 2L and 3L graphene consist of stacked $^{12}C$ and isotopically labelled $^{13}C$ graphene films ($^{13}C\backslash^{12}C$ for 2L and $^{13}C\backslash^{12}C\backslash^{12}C$ for 3L). Inset is schematic of the $SiO_2\backslash^{13}C\backslash^{12}C$ with ads-Ar. (c), Stability of $I_D/I_G$ as prolonging the storage duration. Graphene with different $I_D/I_G$ keep nearly unchanged at RT within 150 days. (d), *in situ* variable-temperature $R/R_{des}$ of graphene films with adsorbed Xe, Kr, Ar, and He with the same ripple deformation. The temperature for desorbing Xe is highest and that for He is lowest. Generally, the $R/R_{des}$ starts to descend from 120 °C and recover to its normal state above 350 °C.

Figure 3d plots the temperature-dependent normalized resistance ($R/R_{des}$) of graphene with adsorbed Xe, Kr, Ar, and He under vacuum ($<10^{-2}$ Pa), where the $R_{des}$ represents the longitudinal resistance ($R_{xx}$) after completely desorption. The desorption processes for all the adsorbed noble gases start at ~120 °C, and the temperatures for completely desorbing Xe, Kr, Ar and He are ~350, ~320, ~250, and ~230 °C, respectively. Notably, despite exhibiting a similar degree of ripple deformation (Fig. S7a), these graphene films with adsorbed Xe, Kr, Ar and He exhibit distinct resistivity. This disparity likely results from differential interactions between adsorbed gas species and the deformed carbon atoms. We further fit the activation energies for gas desorption processes via the Arrhenius law $R = A\cdot\exp(-E_a/k_B T)$, where $R$ is the relative resistance, $k_B$ is the Boltzmann constant, $A$ is pre-exponential factor, $T$ is the measured temperature, and $E_a$ is the activation energy for desorption. The derived values for desorbing Xe, Kr, Ar, and He are 207 meV, 106 meV, 83 meV, and 48 meV, respectively (Fig. S7b-c).

In contrast, all the adsorbed noble gases start to desorb at ~110 °C under AP, which are basically consistent with the desorption under vacuum (Fig. S7d-f). The critical temperatures for complete desorption are independent of the ripple deformation degree for a given noble gas (Fig. S7g), indicating the primary dependence on the adsorbate-adsorbent interaction rather than ripple deformation.

**Adsorption-induced physical property changes**

Figure 4a show the theoretically calculated band structure of rippled graphene with ads-Ar. As the ripple curvature increases, the Fermi level of graphene shifts from -4.25 eV to -4.15 eV, indicating a gradually enhanced *p*-type doping introduced by symmetry breaking. Concurrently, the rippled lattices also induce a bandgap opening of ~20 meV when the ripple curvature reaches 0.116. Additionally, the Fermi level of rippled graphene remains unchanged regardless of the presence of Ar atoms, *i.e.*, the change of band structure mainly results from the rippled lattice rather than adsorbed Ar (Fig. S8).

Figure 4b plots the $R_{xx}$ of as-transferred, ads-Ar and des-Ar graphene films on Si\SiO$_2$ at different gate voltage ($V_g$). The gate voltage at the charge neutrality point ($V_{CNP}$) of as-transferred graphene shifts from 4 to 33.5 V after adsorbing Ar, indicating the *p*-type doping [36,37]. The $V_{CNP}$ for des-Ar graphene recover to 5.5 V. Moreover, the $R_{xx}$ of ads-Ar graphene is nearly two orders of magnitude larger than that of the as-transferred film, which is comparable to hydrogenated and fluorinated graphene with approximate Raman $I_D/I_G$ ratio [33,38], and much larger than graphene with adsorbed polar NH$_3$ or NO$_2$ gases [39]. The $R_{xx}$ and transport characteristic of des-Ar graphene are completely recovered.

Figure 4c further compares the variable-temperature $R_{xx}$ at $V_{CNP}$ ($R_{CNP}$) for as-transferred, ads-Ar and des-Ar graphene, and ads-Ar graphene behave like a typical semiconductor regardless of the different doping levels. The derived transport bandgap of ads-Ar graphene is fitted as ~7.4 meV, indicating the bandgap opening (Fig. S9).

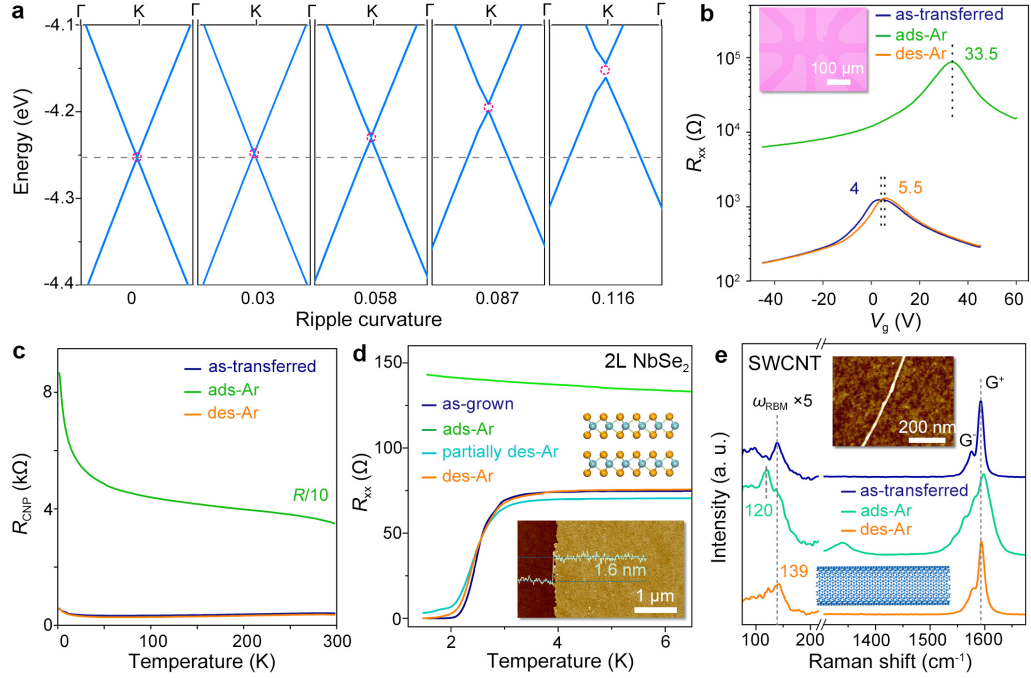

**Figure 4. Physical property changes of 2D materials with adsorbed noble gases.** (a), Theoretical calculated band structures of graphene with different ripple curvature, the curvature from left to right are 0, 0.03, 0.058, 0.087 and 0.116, respectively. (b), Typical electrical transport of as-transferred, ads-Ar, and des-Ar graphene films on Si\SiO$_2$. Inset is the optical image of the graphene Hall bar. (c), Variable-temperature $R_{CNP}$ of as-transferred, ads-Ar, and des-Ar graphene films. (d), Variable-temperature $R_{xx}$ of as-grown, ads-Ar, and des-Ar 2L NbSe$_2$ films. Top inset is the atomic model of 2H phased 2L NbSe$_2$, and bottom inset is the typical AFM image of 2L NbSe$_2$ film on sapphire. (e), Raman spectra of as-transferred, ads-Ar and des-Ar SWCNT. Inset is the corresponding AFM of as-transferred SWCNT on Si\SiO$_2$.

Figure 4d plots the variable-temperature $R_{xx}$ of as-grown, ads-Ar, partially des-Ar and des-Ar NbSe$_2$ films, where the superconductivity for NbSe$_2$ is highly sensitive to the structural change [40]. As-grown 2L NbSe$_2$ films show a superconducting transition critical temperature ($T_c$) of 2.25 K. In contrast, the ads-Ar NbSe$_2$ behaves as a typical semiconductor with nearly doubled $R_{xx}$, and the corresponding Raman characteristic peaks are also changed (Fig. S10a). Partially des-Ar and des-Ar NbSe$_2$ gradually recover the superconducting behaviour with $T_c$ of 1.95 K and 2.15 K, respectively. Similarly, photoluminescence (PL) of 1L MoS$_2$ also changes after Ar adsorption, and the desorption process can also gradually recover the characteristic of semiconductor [41], with slightly changed Raman peaks (Fig. S10b-d).

Individual single-walled carbon nanotube (SWCNT) [42] is also performed, and the Raman radial breathing mode (RBM) of SWCNTs is sensitive to the structural deformation [43]. Figure 4e plot *ex situ* Raman spectra of as-transferred, ads-Ar, and des-Ar individual SWCNT on Si\SiO$_2$ substrate. The as-transferred SWCNT is almost defect-free, and the $\omega_{RBM}$ is located at 139 cm$^{-1}$. After Ar adsorption, the D band appears and additional sub-peaks arise in the G peak region for the same position of SWCNT. More critically, the $\omega_{RBM}$ redshifts to 120 cm$^{-1}$. After desorption, the shapes of D and G peaks along with the $\omega_{RBM}$ all return to the initial state, confirming the recovery of the radial structural deformation.

## CONCLUSIONS

We propose a ripple-assisted adsorption mechanism to realize the stable adsorption of crystallized noble gases on graphene and other layered materials at RT, where the enhanced rippled structure reducing the adsorption energy. The adsorbed noble gas atoms on graphene exhibit packed atomic crystals with periodic arrangement, and the adsorption maintains stable even for half year at RT. They can be desorbed at elevated temperature starting from 110 °C, while the recovered graphene lattice retains completely intact without any structural damages. The elemental characteristics of adsorbed noble gases on graphene are confirmed comprehensively. The adsorption-induced ripple deformation modifies the structural and physical properties of graphene, NbSe$_2$, MoS$_2$, and SWCNT, where their electrical transport, superconductivity, PL, and Raman RBM are hugely affected. These property changes can be completely recovered after complete desorption. The ripple-assisted adsorption is totally different from conventional physisorption and chemisorption theory, and hence we believe this adsorption for different gases, especially nonpolar gases, will accelerate the process of gas storage and separation technologies, as well as facilitate their applications in catalysis, surface modification, and other related fields.

## SUPPLEMENTARY DATA

Supplementary data are available at *NSR* online.

## FUNDING

This work is financially supported by the National Natural Science Foundation of China (NSFC, grants 52425203, 12404216, 12374183, 92165205, T2321002, 22173019), Natural Science Foundation of Jiangsu Province (grants BK20240008, BK20250063, BK20241252, BK20233001, BK20222007), National Key Research and Development Program of China (grant 2021YFA1400403), Xiaomi

Foundation, Fund of laboratory of Solid State Microstructures of Nanjing University (grant no. M35041), the program A for outstanding PhD candidate of Nanjing University. Postdoctoral Fellowship Program of CPSF (grants GZC20231093), Jiangsu Funding Program for Excellent Postdoctoral Talent (grant 2023ZB553), the Innovation Program for Quantum Science and Technology (grant 2021ZD0302800) and the Fundamental Research Funds for the Central Universities (grant 020414380207). Q. F., J. W. and M. J. thank the Big Data Computing Center of Southeast University for providing the facility support on the calculations.

**AUTHOR CONTRIBUTIONS**

L.G. and G.Y. supervised the project and conceived the experiments. W.L. and X.H. performed gas adsorption, OM, Raman, AFM and electrical measurements. G.Y. assisted in gas adsorption and desorption, graphene growth and transfer, AFM, electrical measurements. L.D., Y.X. and S.L. performed STM and STS measurements. Q.F., J.W. and M.J. performed the theoretical calculation. Ang Li and W.Z. performed EELS and SAED measurements. Z.Z, J.L., Y.J., Z.H. and Z.F assisted in the growth of $NbSe_2$ and $MoS_2$. P.H. and C.L. provided SWCNTs. L.G., G.Y., W.L. and X.H. analysed the data and wrote the manuscript, W.Z., M.J., S.L., and H.M.C. revised it, and all authors commented on it.

*Conflict of interest statement.* None declared

**REFERENCE**

1. Freundlich HM. Over the Adsorption in Solution. *Z Phys Chem* 1906; **57**: 385-470.
2. Langmuir I. The constitution and fundamental properties of solids and liquids Part I Solids. *J Am Chem Soc* 1916; **38**: 2221-95.
3. Langmuir I. The constitution and fundamental properties of solids and liquids. II. Liquids. *J Am Chem Soc* 1917; **39**: 1848-906.
4. Brunauer S, Emmett PH, Teller E. Adsorption of gases in multimolecular layers. *J Am Chem Soc* 1938; **60**: 309-19.
5. Kim M, Bertram M, Pollmann M *et al.* Controlling chemical turbulence by global delayed feedback: Pattern formation in catalytic CO oxidation on Pt(110). *Science* 2001; **292**: 1357-60.
6. Bassignana IC, Wagemann K, Kuppers J *et al.* Adsorption and Thermal-Decomposition of Ammonia on a Ni(110) Surface : Isolation and Identification of Adsorbed $NH_2$ and NH. *Surf Sci* 1986; **175**: 22-44.
7. Low MJD. Kinetics of Chemisorption of Gases on Solids. *Chem Rev* 1960; **60**: 267-312.
8. Huber F, Berwanger J, Polesya S *et al.* Chemical bond formation showing a transition from physisorption

to chemisorption. *Science* 2019; **366**: 235-8.
9. Morris RE, Wheatley PS. Gas storage in nanoporous materials. *Angew Chem Int Edit* 2008; **47**: 4966-81.
10. Allendorf MD, Stavila V, Snider JL *et al.* Challenges to developing materials for the transport and storage of hydrogen. *Nat Chem* 2022; **14**: 1214-23.
11. Subrahmanyam KS, Kumar P, Maitra U *et al.* Chemical storage of hydrogen in few-layer graphene. *P Natl Acad Sci USA* 2011; **108**: 2674-7.
12. Sun PZ, Yagmurcukardes M, Zhang R *et al.* Exponentially selective molecular sieving through angstrom pores. *Nat Commun* 2021; **12**: 7170.
13. Christmann K, Behm RJ, Ertl G *et al.* Chemisorption Geometry of Hydrogen on Ni(111) : Order and Disorder. *J Chem Phys* 1979; **70**: 4168-84.
14. Freund HJ. Adsorption of Gases on Complex Solid surfaces. *Angew Chem Int Edit* 1997; **36**: 452-75.
15. Park JY, Kahng SJ, Ham UD *et al.* Adsorption and growth of Xe adlayers on the Cu(111) surface. *Phys Rev B* 1999; **60**: 16934-40.
16. Li YJ, Takeuchi O, Futaba DN *et al.* Characteristic adsorption of Xe on a Si(111)-(7×7) surface at low temperature. *Phys Rev B* 2002; **65**: 113306.
17. Li YJ, Takeuchi O, Futaba DN *et al.* Characteristic intra- and interunit interactions of Kr atoms adsorbed on the Si(111)-7×7 surface. *Phys Rev B* 2003; **68**: 033301.
18. Li JR, Kuppler RJ, Zhou HC. Selective gas adsorption and separation in metal-organic frameworks. *Chem Soc Rev* 2009; **38**: 1477-504.
19. Cohen PI, Unguris J, Webb MB. Xe Monolayer Adsorption on Ag(111) .1. Structural-Properties. *Surf Sci* 1976; **58**: 429-56.
20. Eigler DM, Weiss PS, Schweizer EK *et al.* Imaging Xe with a Low-Temperature Scanning Tunneling Microscope. *Phys Rev Lett* 1991; **66**: 1189-92.
21. Bauerle C, Mori N, Kurata G *et al.* Studies of 2D cryocrystals by STM techniques. *J Low Temp Phys* 1998; **113**: 927-32.
22. Adams MA, Mayers J, Kirichek O *et al.* Measurement of the kinetic energy and lattice constant in hcp solid helium at temperatures 0.07-0.4 K. *Phys Rev Lett* 2007; **98**: 085301.
23. Konig T, Simon GH, Rieder KH *et al.* Superlattice structure of an Ar monolayer on Ag(111) observed by low-temperature scanning tunneling microscopy. *Phys Rev B* 2008; **78**: 201407.
24. Kawai S, Foster AS, Björkman T *et al.* Van der Waals interactions and the limits of isolated atom models at interfaces. *Nat Commun* 2016; **7**: 11559.
25. Fasolino A, Los JH, Katsnelson MI. Intrinsic ripples in graphene. *Nat Mater* 2007; **6**: 858-61.
26. Meyer JC, Geim AK, Katsnelson MI *et al.* The structure of suspended graphene sheets. *Nature* 2007; **446**: 60-3.
27. Sun PZ, Xiong WQ, Bera A *et al.* Unexpected catalytic activity of nanorippled graphene. *P Natl Acad Sci USA* 2023; **120**: e2300481120.
28. Xiong WQ, Zhou WQ, Sun PZ *et al.* Enhanced hydrogen-gas permeation through rippled graphene. *Phys Rev B* 2023; **108**: 045408.
29. Sun PZ, Yang Q, Kuang WJ *et al.* Limits on gas impermeability of graphene. *Nature* 2020; **579**: 229-32.
30. Andreev T, Barke I, Hövel H. Adsorbed rare-gas layers on Au(111): Shift of the Shockley surface state studied with ultraviolet photoelectron spectroscopy and scanning tunneling spectroscopy. *Phys Rev B* 2004; **70**: 205426.
31. Ferrari AC, Basko DM. Raman spectroscopy as a versatile tool for studying the properties of graphene. *Nat Nanotechnol* 2013; **8**: 235-46.

32. Lee JE, Ahn G, Shim J *et al.* Optical separation of mechanical strain from charge doping in graphene. *Nat Commun* 2012; **3**: 1024.
33. Elias DC, Nair RR, Mohiuddin TMG *et al.* Control of Graphene's Properties by Reversible Hydrogenation: Evidence for Graphane. *Science* 2009; **323**: 610-613.
34. Huang XL, Zheng H, Liu WL *et al.* Realization of highly asymmetric hydrogenated graphene in the van der Waals confined space. *Natl Sci Rev* 2025; **12**: nwaf067.
35. Yoon K, Rahnamoun A, Swett JL *et al.* Atomistic-Scale Simulations of Defect Formation in Graphene under Noble Gas Ion Irradiation. *ACS Nano* 2016; **10**: 8376-84.
36. Yuan GW, Lin DJ, Wang Y *et al.* Proton-assisted growth of ultra-flat graphene films. *Nature* 2020; **577**: 204-8.
37. Yuan GW, Liu WL, Huang XL *et al.* Stacking transfer of wafer-scale graphene-based van der Waals superlattices. *Nat Commun* 2023; **14**: 5457.
38. Robinson JT, Burgess JS, Junkermeier CE *et al.* Properties of Fluorinated Graphene Films. *Nano Lett* 2010; **10**: 3001-5.
39. Schedin F, Geim AK, Morozov SV *et al.* Detection of individual gas molecules adsorbed on graphene. *Nat Mater* 2007; **6**: 652-5.
40. Lin HH, Zhu Q, Shu DH *et al.* Growth of environmentally stable transition metal selenide films. *Nat Mater* 2019; **18**: 602-7.
41. Zhu JQ, Wang ZC, Yu H *et al.* Argon Plasma Induced Phase Transition in Monolayer $MoS_2$. *J Am Chem Soc* 2017; **139**: 10216-9.
42. Jiang S, Hou PX, Chen ML *et al.* Ultrahigh-performance transparent conductive films of carbon-welded isolated single-wall carbon nanotubes. *Sci Adv* 2018; **4**: eaap9264.
43. Chiashi S, Kono K, Matsumoto D *et al.* Adsorption effects on radial breathing mode of single-walled carbon nanotubes. *Phys Rev B* 2015; **91**: 155415.

# Supporting Information for

# Ripple-assisted adsorption of noble gases on graphene at room temperature

Weilin Liu[1,†], Xianlei Huang[1,†], Li-Guo Dou[1,†], Qianglong Fang[2,†], Ang Li[3,†], Guowen Yuan[1,*], Yongjie Xu[1], Zhenjia Zhou[1], Jun Li[1], Yu Jiang[1], Zichong Huang[1], Zihao Fu[1], Peng-Xiang Hou[4], Chang Liu[4], Jinlan Wang[2,5], Wu Zhou[3*], Ming-Gang Ju[2,*], Shao-Chun Li[1,6,*], Hui-Ming Cheng[4,7] and Libo Gao[1,8,*]

[1] National Laboratory of Solid State Microstructures, Jiangsu Key Laboratory for Nanotechnology, Jiangsu Physical Science Research Center, School of Physics, Nanjing University, Nanjing, China.

[2] Key Laboratory of Quantum Materials and Devices of Ministry of Education, School of Physics, Southeast University, Nanjing, China.

[3] School of Physical Sciences, University of Chinese Academy of Sciences, Beijing, China.

[4] Shenyang National Laboratory for Materials Sciences, Institute of Metal Research, Chinese Academy of Sciences, Shenyang, China.

[5] Suzhou Laboratory, Suzhou, China.

[6] Hefei National Laboratory, Hefei, China.

[7] Institute of Technology for Carbon Neutrality, Shenzhen Institute of Advanced Technology, Chinese Academy of Sciences, Shenzhen, China.

[8] State Key Laboratory of Chemo/Biosensing and Chemometrics, Key Laboratory for Micro-Nano Physics and Technology of Hunan Province, College of Materials Science and Engineering, Hunan University, Changsha, China.

*Corresponding author. E-mail: gwyuan@nju.edu.cn; wuzhou@ucas.ac.cn; juming@seu.edu.cn; scli@nju.edu.cn; lbgao@nju.edu.cn

†Equally contributed to this work.

# EXPERIMENTAL AND METHODS

**Sample preparation**

**Exfoliated graphene flakes.** Few-layer graphene flakes are exfoliated onto Si substrates with 285 nm thick oxide (Si\SiO$_2$) by a traditional micromechanical cleavage method from kish graphite. Then, all the flakes are annealed under high vacuum (<10$^{-5}$ Pa) at 350 °C for 30 min to remove the polymer residues.

**CVD graphene films.** Large-area ultra-flat graphene films are grown by the proton-assisted chemical vapour deposition method as reported previously[35]. The typical growth parameters are as follows: sputtered 800 nm thick Cu-Ni(111) films on c-plane sapphire as substrates, growth temperature of 650 °C, pressure of 6 Pa, CH$_4$/H$_2$ ratio of 1:20, plasma power of 15 W and growth time of 5 min. The growth parameters for traditional CVD grown graphene films are as follows: sputtered 800 nm thick Cu-Ni(111) films on c-plane sapphire as substrates, growth temperature of 1050 °C, CH$_4$/H$_2$/Ar ratio of 0.1:10:500 under AP, growth time of 10 min. We use Cu-Ni alloy (90% Cu + 10% Ni) instead of Cu in this study to avoid sublimation at 1050 °C or vacuum annealing at 650 °C. To simplify the writing, we use Cu(111) instead of Cu-Ni(111) alloy in the main text.

**CVD NbSe$_2$ films.** 2L NbSe$_2$ are grown by the two-step vapour deposition method as reported previously. The typical sputter parameters are as follows: totally 1.0 nm thick Nb films, c-plane sapphire as sputtered substrate, substrate temperature of ~200 °C, deposition rate of ~0.2 Å/s and the constant chamber pressure of 10 Pa. The typical CVD growth parameters are as follows: heating Se powder (>99.9%) of 340 °C in zone I upstream, heating Nb films of 650 °C in zone II downstream, growth time of 15 min, the carrier gas of H$_2$/Ar (100:100) under AP. After the growth, the NbSe$_2$ films are annealed at 400 °C for 30 min in zone II to remove the redundant Se particles (zone I is not heated).

**CVD MoS$_2$ grains.** 1L MoS$_2$ grains are grown by the two-step vapour deposition as reported previously[43]. The sputter and CVD growth parameters are similar to those of NbSe$_2$ above. The sputtered Mo films are 0.7 nm, and the temperatures of heating S powder (>99.9%) and Mo film are 160 °C and 800 °C, respectively.

**CVD SWCNT.** An injection floating catalyst chemical vapor deposition (FC-CVD) is used for growing the isolated SWCNT networks as reported previously[41]. The growth and precursor temperatures are 1100 and 83 °C with pure Ar atmosphere under AP, respectively. Then, the carrier gas of $C_2H_4/H_2$ ratio of 5:2 is introduced, along with the mixed solution containing 96 wt.% toluene, 3 wt.% ferrocene, and 1 wt.% thiophene injected into the reactor at a rate of 4 sccm through a syringe pump. The SWCNT films with different thicknesses are collected on porous cellulose filter membranes installed at the outlet of the flowing gases. The individual SWCNTs are transferred from the porous cellulose filter membranes onto the Si\SiO$_2$ substrates by simple pressing.

**Transfer of CVD graphene films.** After CVD growth, the sapphire\Cu(111)\graphene is spin-coated with double layers of polymethyl methacrylate (PMMA, first layer, 120k MW, 1 wt.% in ethyl lactate, 2000 rpm for 1 min; second layer, 996k MW, 4 wt.% in ethyl lactate, 2000 rpm for 1 min) and baked at 150 °C in air for 10 min. Then, 1 M $(NH_4)_2S_2O_8$ aqueous solution is used to etch the Cu(111). After cleaned with DI water thrice, the floating graphene\PMMA is pasted onto the target substrate, including Si\SiO$_2$ wafers and microgrids. Subsequently, substrate\graphene\PMMA is baked at 40 °C for 6 h, 80 °C for 30 min and 150 °C for 10 min in sequence. Finally, PMMA films are removed by acetone and the further proton-assisted cleaning process[36].

## Characterizations

**Optical microscope.** All the optical images are captured by the optical microscope (Nikon Eclipse LV100ND).

**Raman.** Raman spectra are acquired using a WITec/alpha 300R confocal microscope with 532 nm laser under ambient conditions. The variable-temperature Raman measurements are performed in a temperature-controlled stage (Linkam THMS600), which is integrated in the Raman system. All the laser power is set below 2 mW to avoid heating.

**AFM.** AFM measurements are performed with a Bruker Dimension Fastscan system at tapping mode.

**XPS.** XPS is performed using a commercial PHI 5000 X-ray photoelectron spectrometer equipped with a monochromatized Al Kα radiation ($hv$ = 1486.6 eV). The measurements are performed at RT under UHV with the X-ray power of 25 W. The X-ray spot size is approximately 100 × 100 μm$^2$, while the sampling analysis area is about 1350 × 700 μm$^2$. Peak fitting is performed using a mixed Lorentzian-Gaussian function.

**STM and STS.** The STM measurements of ads-Xe, ads-Ar, and ads-He on Cu(111)\Gra are performed under UHV at 77 K (USM 1500, Unisoku). The topographic images are measured under the constant current mode, and the differential conductance (d$I$/d$V$) spectra are taken using a standard lock-in techniques at 77 K ($f$=879 Hz, $\Delta V_{rms}$=12 mV). Partially desorption process is performed to obtain the clear high-resolution imaging. The ads-Xe, ads-Ar, ads-He graphene in Fig. 1c, 1i and 1j are annealed under UHV at 90 °C for 1.5 h, 180 °C for 1.5 h, 140 °C for 12 h before STM measurement, respectively.

**STEM.** The STEM measurements are performed on a Nion U-HERMES100 microscope with an acceleration voltage of 60 kV under UHV. The convergence semi-angle is set to 32 mrad, and the e-beam current for imaging during the experiments is ~18 pA. The STEM-EELS measurements are performed using a collection semi-angle of 75 mrad, an energy dispersion of 0.9 eV per channel and a probe current of ~0.1 pA. The dose rate is further reduced by a large defocus probe (~5 μm). The EELS camera is a direct electron EELS detector (DECTRIS ELA) with single electron sensitivity, enabling the low dose rate measurement. The SAED patterns are collected in a JEOL 2100Plus TEM with an acceleration voltage of 80 kV.

**Electrical properties.** The electrical transport measurements and variable-temperature resistance measurements of graphene, and NbSe$_2$ films below 300 K are performed in a $^4$He cryostat with a superconducting magnet (Oxford Teslatron 8 T), and the standard four-terminal method is adopted with a lock-in amplifier (Stanford SR830) at a frequency of 3.67 Hz and the bias current of 100 nA. The variable-temperature resistance measurements above 300 K of graphene films are performed in a heating stage under vacuum (<10$^{-2}$ Pa) with the bias current of 1 μA (Keithley 2450) and monitored voltage (Keithley 2182A). For superconductivity, the $T_c$ temperature is calibrated as the sheet resistance drops to 10% of its normal state.

**Controllable adsorption of gas molecules**

Three methods are used to inject net electrons in graphene. The specific operations are as follows:

**Charge injection in SEM.** The negative charges (electron) are controllably emitted from electron gun of environmental scanning electron microscope (SEM, FEI Quanta 200). To avoid high-energy electrons irradiate graphene directly, electrons are chosen to radiate the Cu(111) region of Cu(111)\Gra sample and soldered indium of SiO$_2$\Gra samples. The Ar gas flow into SEM by a valve. Typical adsorption parameters are as follows: constant pressure of 10 Pa, accelerating voltage of 30 kV, emission current of ~80 pA, adsorption time of 30 min, at RT.

**Charge injection in customized instrument.** Customized instrument is designed to mimic the charge injection process in SEM. The negative charges (electron) are controllably emitted from an electrostatic generator (JTStar, JT206B), which consists of several tungsten tips and a high voltage power supply (0 to -20 kV). The emitted electrons are collected at a copper plate and transmitted to the target samples via a copper wire. The injected electron density is monitored by a high precision ampere meter (KangWei, KV-MCM02503) calibrated by a source meter (Keithley 6430). To avoid the glow discharge, the charging density by election injection is usually <10 nA/mm$^2$. The typical adsorption parameters are as follows: gas pressure of 133 Pa (1 Torr), charging current density of 0.55 nA/mm$^2$. The adsorption degree is modified by the charging time.

**Plasma treatment.** The controllable plasma of different gas is generated in a customized inductively coupled plasma (ICP) system, and typical treatment parameters are as follows: flow rate of 120 sccm, constant pressure of 6 Pa, plasma power of 10 W, treatment time of 10 – 120 s, adsorption temperature of 35 °C.

**Desorption by thermal annealing.** For *ex situ* desorption, all the treatments are performed in a high vacuum (<10$^{-5}$ Pa), and their thermal annealing conditions are as follows: graphene, 350 °C, 30 min; MoS$_2$, 350 °C, 30 min; NbSe$_2$, 150 °C, 30 min for partial desorption, 170 °C, 30 min for complete desorption; SWCNT, 350 °C, 30 min.

**Theoretical calculation for total adsorption energy**

A graphene sheet with a diameter of 12.4 Å passivated by hydrogen atoms is chosen to model the adsorption of noble gas on the rippled graphene surface. Geometric optimization and energy calculations are performed using the Gaussian 16 program package at the M06-2X/SDD level of theory for noble gases, including He, Ne, Ar, Kr and Xe. This level of theory is selected to fully consider the relativistic effects. The atoms in the upper layer and the bottom layer are fixed, while the atoms in the middle layer relax freely. The adsorption energy ($E_{ads}$) is defined as:

$$E_{ads} = E_{Gra+gas} - E_{Gra} - E_{gas}$$

Where $E_{Gra+gas}$, $E_{Gra}$ and $E_{gas}$ are the total energies of rippled graphene adsorbed with noble gas, single ripple graphene sheet and one noble gas, respectively.

The band structures of graphene with different ripple curvature are calculated based on the density functional theory (DFT) implemented in the Vienna Ab initio Simulation Package (VASP) with the projector-augmented wave method. The GGA-PBE function is used to describe the exchange-correlation functionals. The plane-wave cutoff energy is set to 450 eV, and the force converged to 0.01 eV/Å. Dipole correction is employed to eliminate the potential errors, and the van der Waals interaction between graphene and Ar atom is included by DFT-D3.

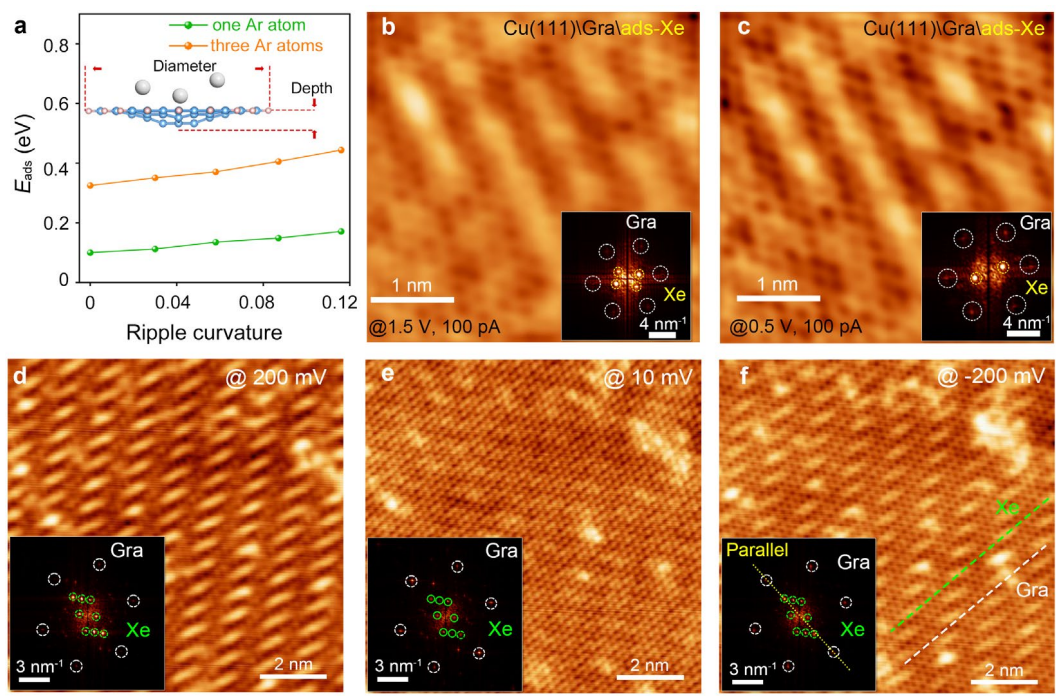

**Figure S1. Additional STM images of ads-Xe graphene on Cu(111)**. **a** Theoretical calculated $E_{ads}$ of graphene with different ripple curvature adsorbing noble gas atoms. Inset illustrates the side view of rippled graphene (96 carbon atoms) adsorbing three Ar atoms. **b-c**, *in situ* STM images of ads-Xe graphene measured at different bias voltage share the same position as those in main Fig. 1d and 1e. Insets are the corresponding FFT patterns. **d-f**, *in situ* STM images of ads-Xe graphene measured with different bias voltage. As changing the bias, the surface morphologies show obviously different, where the Xe atomic crystals are clearer when applying higher bias voltage and graphene lattice becomes clearer when applying lower bias voltage. Insets are the corresponding FFT patterns. The Xe atomic crystal arrangements exhibit a consistent orientation with the graphene lattice's crystalline structure, as indicated by the dashed line in panel (**f**).

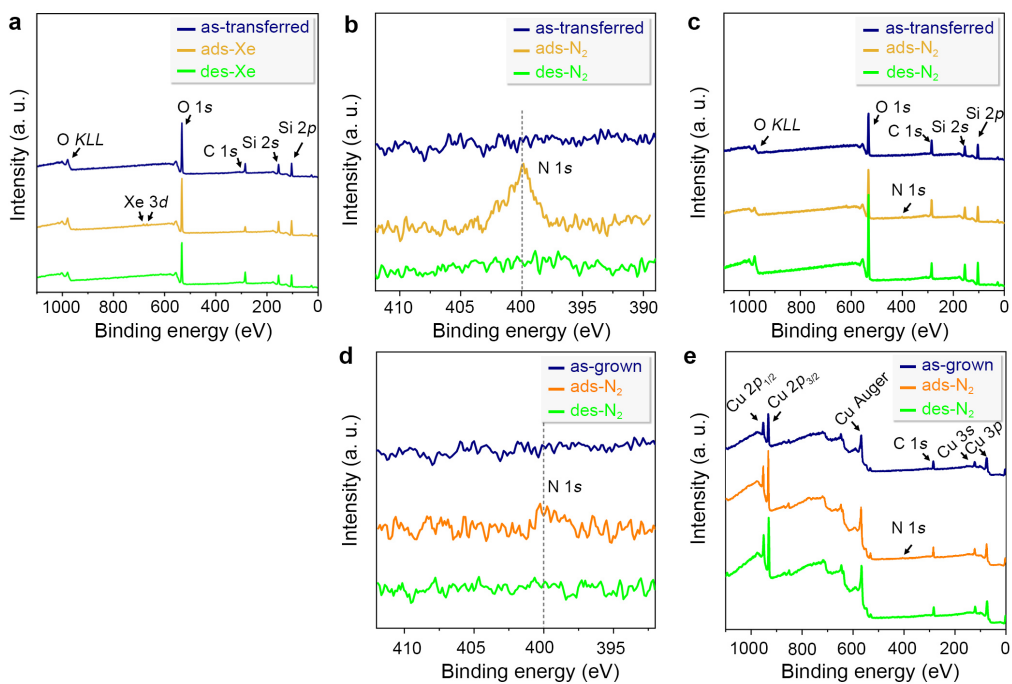

**Figure S2. Additional XPS data for ads-Xe and ads-N$_2$ graphene**. **a**, Corresponding survey XPS spectra of as-transferred, ads-Xe and des-Xe graphene on Si\SiO$_2$. **b**, Fine XPS spectra of as-transferred graphene, ads-N$_2$ graphene and des-N$_2$ graphene (vacuum annealing at 300 °C) on Si\SiO$_2$. The apparent N characteristic of N-N core level is located at ~400 eV for ads-N$_2$ graphene. **c**, Corresponding survey XPS spectra of as-transferred graphene, ads-N$_2$ and des-N$_2$ graphene on Si\SiO$_2$. **d**, Fine XPS spectra of as-grown, ads-N$_2$ and des-N$_2$ graphene on Cu(111) under the same adsorption condition with **b**. The apparent N characteristic of N-N core level is located at ~400 eV for ads-N$_2$ graphene. **f**, Corresponding survey XPS spectra of as-grown, ads-N$_2$ and des-N$_2$ graphene on Cu(111).

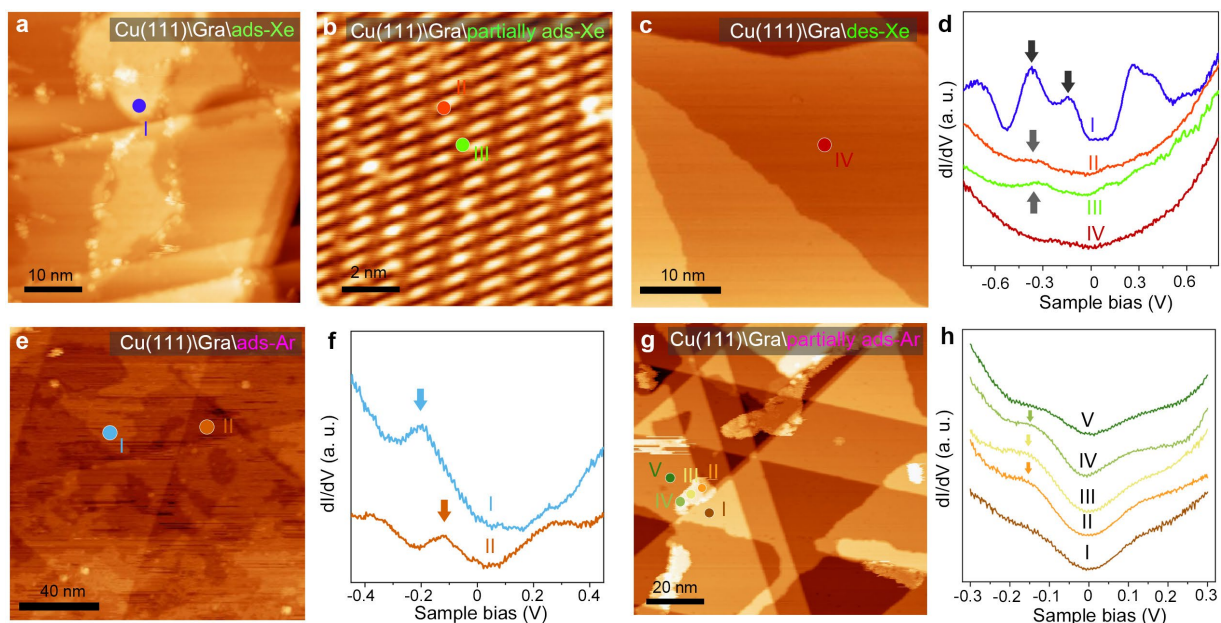

**Figure S3. Additional STM images and STS spectra of ads-Xe and ads-Ar graphene on Cu(111)**. **a**, Typical STM topological image of ads-Xe graphene. **b**, Typical STM topological image of partially ads-Xe graphene after 180 °C annealing, part of adsorbed Xe remains. **c**, Typical STM topological image of des0Xe graphene after 350 °C annealing, no adsorbed Xe left. **d**, Corresponding STS spectra from the labelled regions in **a-c**. There are shows shoulder peaks (arrow-marked) and a bandgap opening for ads-Xe graphene. **e**, Typical STM topological image of ads-Ar graphene. **f**, Corresponding STS spectra from the labelled regions in **e**. There are shows shoulder peaks (arrow-marked) and a bandgap opening for ads-Ar graphene. **g**, Typical STM topological image of partially ads-Ar graphene after 180 °C annealing, part of adsorbed Ar remains. **h**, Corresponding STS spectra from the labelled regions in **g**. The shoulder peak persists in ads-Ar graphene with the changed peak position and intensity.

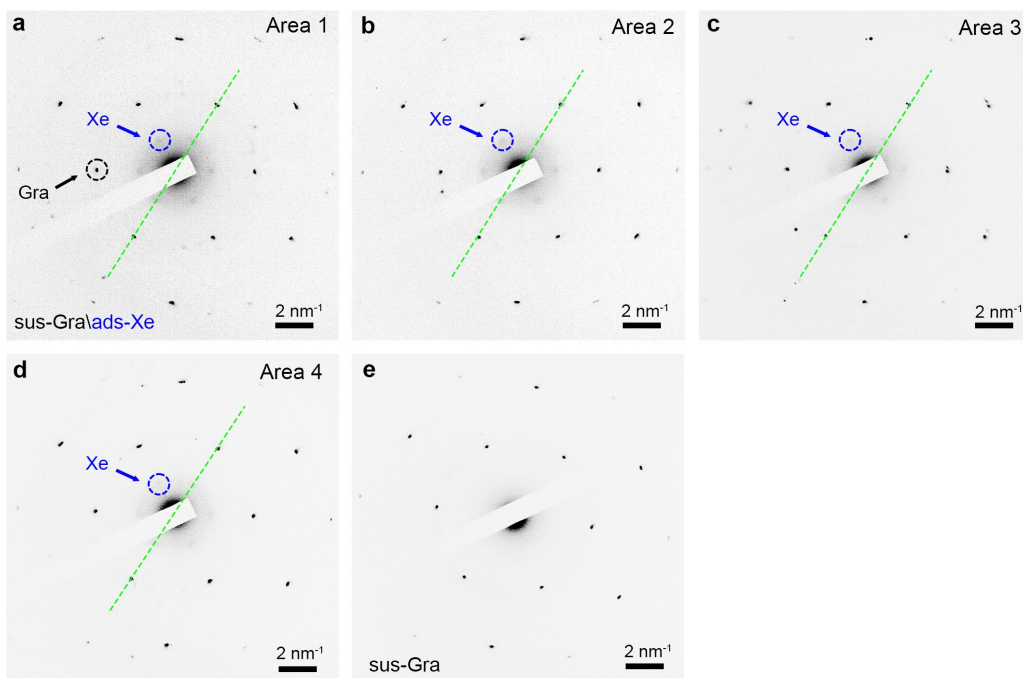

**Figure S4. Additional SAED patterns of ads-Xe on suspending graphene**. **a-d**, Additional SAED patterns of graphene with adsorbed Xe atoms at different locations. The black and blue dashed circles labelled in diffraction patterns represent graphene and crystallized Xe, respectively. The dashed lines indicate that the diffraction spots of crystallized Xe always align with the diffraction spots of graphene. **e**, SAED pattern of pure suspending graphene.

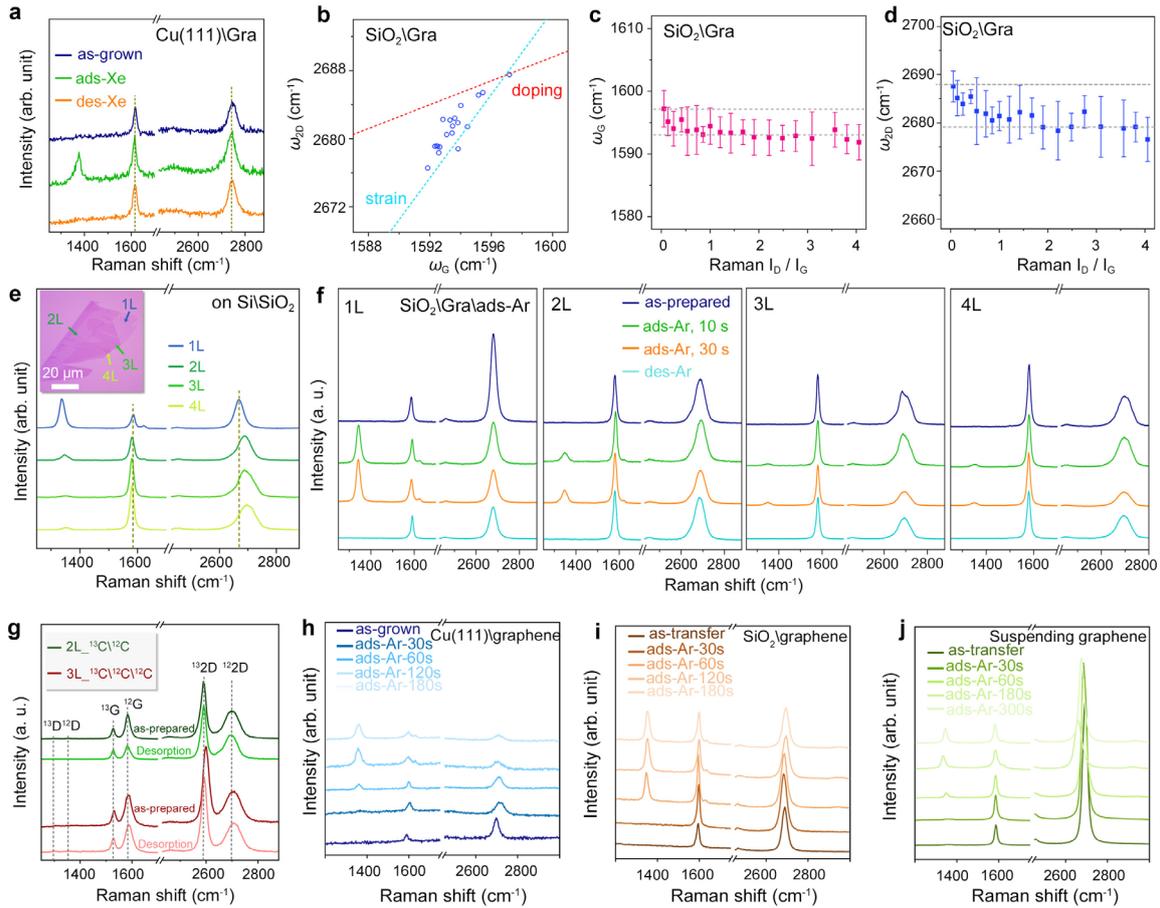

**Figure S5. Additional Raman data of ads-Xe and ads-Ar graphene**. **a**, Raman spectra of as-grown, ads-Xe and des-Xe graphene on Cu(111). The adsorbed Xe atoms induce the *sp*$^3$ phase transition (D peak) and the rippled deformation of graphene can be completely recovered. **b**, Relationship between $\omega_G$ and $\omega_{2D}$, the dash lines are guides to the strain and doping induced frequency shifts, indicating the tension strains are introduced when graphene adsorbing Ar. **c**, Relationship between the frequency of G peak ($\omega_G$) and $I_D/I_G$ for ads-Ar graphene, the $\omega_G$ shows slight redshift of ~5 cm$^{-1}$. **d**, Relationship between the frequency of 2D band ($\omega_{2D}$) and $I_D/I_G$ for ads-Ar graphene, the $\omega_{2D}$ shows slight redshift of ~8 cm$^{-1}$. **e**, Raman spectra of graphene with different layer number adsorbed by Ar atoms. Inset is the corresponding optical image of exfoliated graphene flakes on Si\SiO$_2$. **f**, Raman spectra of exfoliated graphene varying layer numbers on Si\SiO$_2$ after adsorbing Ar through weak ICP with different adsorption time, from left to right is 1L, 2L, 3L and 4L. The $I_D/I_G$ are rapidly reduced as increasing the layer number of graphene, indicating the ripple deformation is restricted for thick layer structures. **g**, Raman spectra of as-transferred and des-Ar 2L & 3L graphene stacked by the top $^{12}$C and bottom isotopic $^{13}$C. The rippled deformation of graphene can be completely recovered. **h-j.** Raman spectra of graphene on Cu(111), Si\SiO$_2$ and suspending graphene after adsorbing Ar through weak ICP with different adsorption time. As the adsorption time increasing, the change rate of the $I_D/I_G$ of graphene on Cu(111) is the fastest, while that of the suspended graphene film is the slowest.

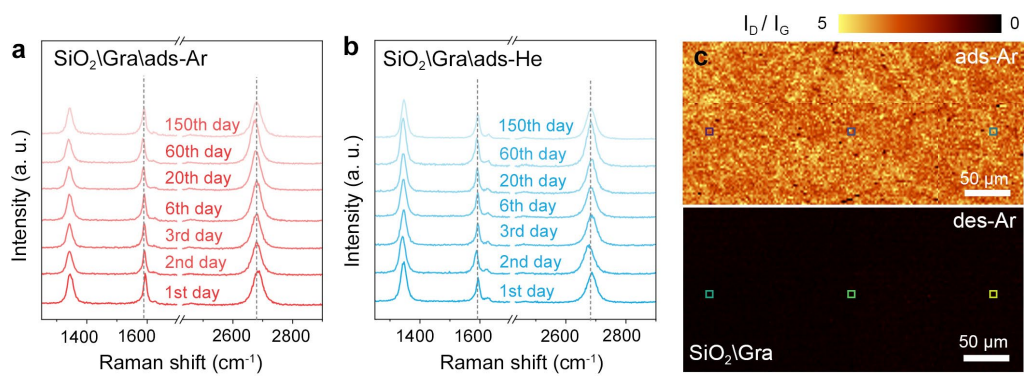

**Figure S6. Additional Raman data of adsorption stability and uniformity**. **a-b**, Typical Raman spectra of ads-Ar graphene stored for different durations under vacuum, these data correspond to the main Fig. 3c, indicating the adsorption are very stable at RT. **c**, Raman $I_D/I_G$ mapping of ads-Ar and des-Ar graphene films. The $I_D/I_G$ are both homogenous across the large area.

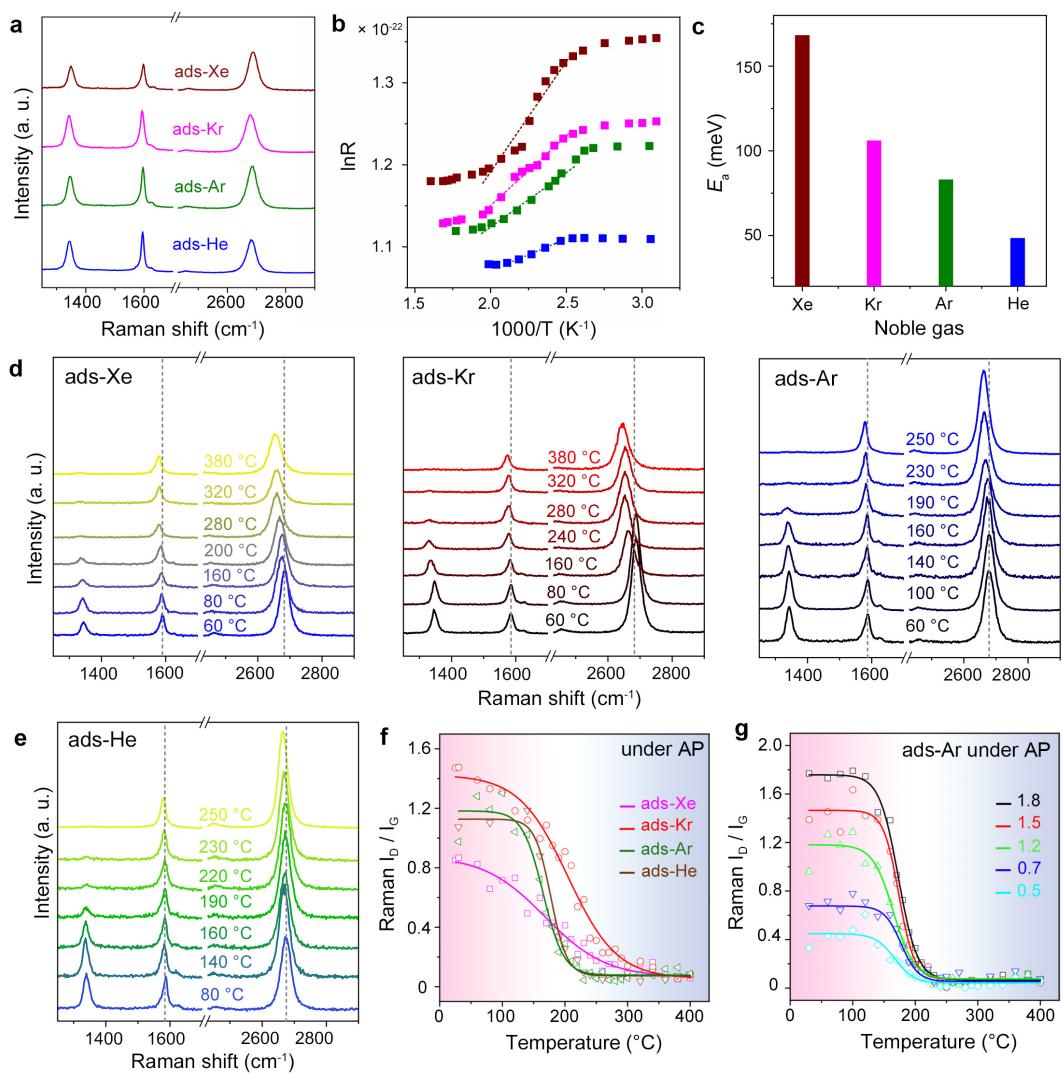

**Figure S7. Additional Raman data of ads-gas or des-gas graphene on Si\SiO$_2$**. **a**, Corresponding Raman spectra of ads-Xe, ads-Kr, ads-Ar and ads-He graphene in the main Fig. 3d, they have the similar I$_D$/I$_G$ but the different resistances. **b**, The resistance of gas-adsorbed graphene at different annealing temperatures. Coloured dash lines are used to denote the fitted slopes in regions of significant resistance variation. **c**, Corresponding activation energies ($E_a$) for gas desorption from graphene, as derived from the data in (**b**). **d-e**, *in situ* Raman spectra of ads-Xe, ads-Kr, ads-Ar and ads-He graphene films as elevating temperature, all their D peaks completely disappear at above 350 °C. **f**, Extracted Raman I$_D$/I$_G$ evolution of ads-Xe, ads-Kr, ads-Ar and ads-He graphene films as elevating temperature under AP. Their non-*sp*$^2$ phased lattices start to be reduced from 100–150 °C, and all the lattices can be completely recovered above 350 °C. The noble gas with larger atomic weight has a higher temperature for complete desorption. **g**, Extracted Raman I$_D$/I$_G$ evolution of ads-Ar graphene with different ripple deformations as elevating temperature under AP. Although the ripple deformations are different, the desorption starting and finishing temperature keep the consistent.

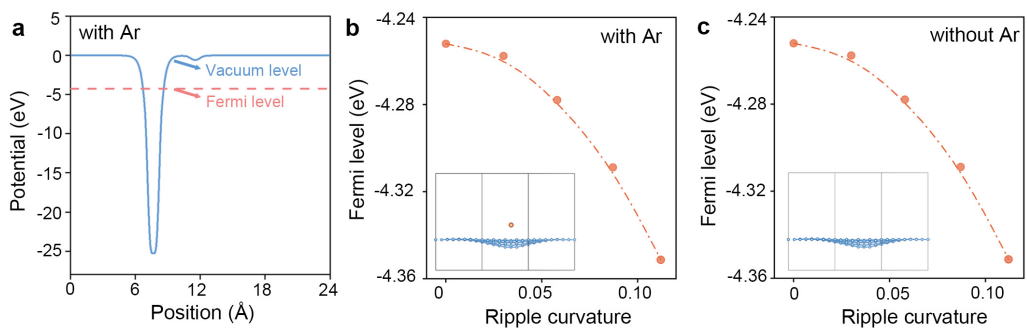

**Figure S8. Additional data for calculated band structure of rippled graphene**. **a**, Theoretical calculated potential energy of rippled graphene, where the vacuum level is shown as reference and initial value is set to 0. **b**, Calculated Fermi level of rippled graphene with adsorbed Ar as the function of ripple curvature, inset is the typical side view of one adsorbed Ar atom on rippled graphene. **c**, Calculated Fermi level of pure graphene as the function of ripple curvature, inset is the typical side view of rippled graphene.

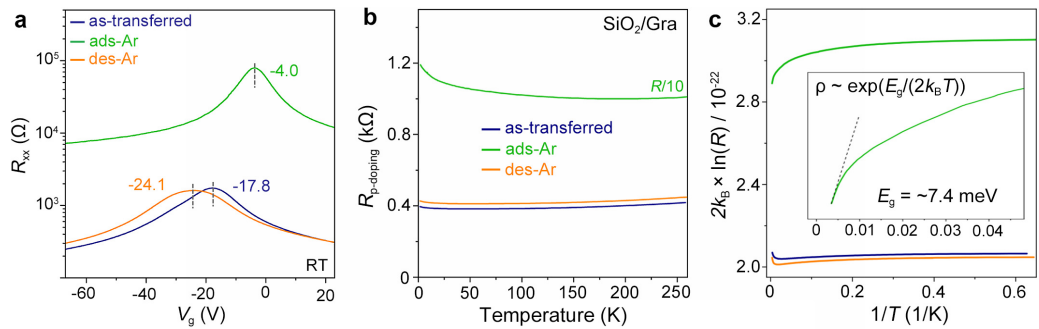

**Figure S9. Additional data for variable-temperature resistance of ads-Ar graphene**. **a**, The electrical transport of as-transferred, ads-Ar, and des-Ar graphene films on Si\SiO$_2$. **b**, Variable-temperature resistances of as-transferred, ads-Ar and des-Ar graphene films on Si\SiO$_2$, the resistances are collected at the high doping density. The resistances of ads-Ar graphene are divided by tenfold for better comparison. The doping level of graphene is tuned by the gate voltage, the doping density of graphene near CNP is the lowest, and the high doping density is far away from CNP. Here, we use *p*-type doping of +30 V from CNP. **c**, Calculated activation energy of as-transferred, ads-Ar and des-Ar graphene on Si\SiO$_2$, the resistances are collected at CNP and converted from the Fig. S9a. Inset is the fitting for ads-Ar graphene, and the band gap is estimated to be 7.4 meV.

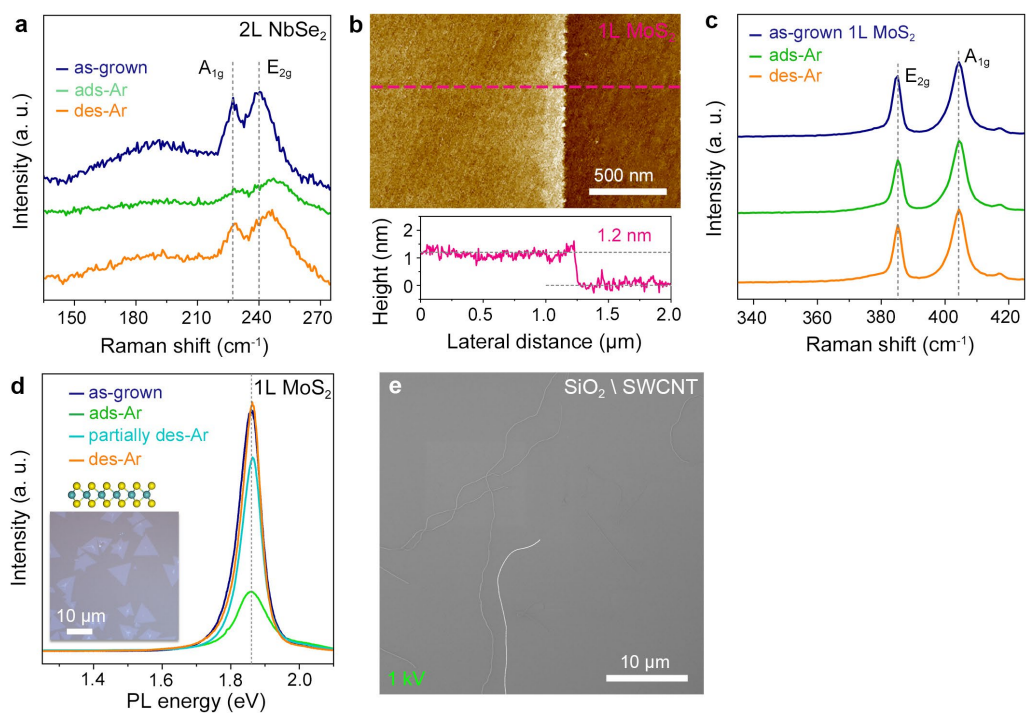

**Figure S10. Additional data for stable adsorption of Ar atoms on more low-dimensional materials.**
**a**, Typical Raman spectra of as-grown, ads-Ar and des-Ar 2L NbSe$_2$ on sapphire. The characteristic peaks of ads-Ar NbSe$_2$ are blueshift and the intensities are reduced. The characteristic peaks of des-Ar NbSe$_2$ can be completely recovered. **b**, Typical AFM image of 1L MoS$_2$ film on sapphire. They are homogenous with the thickness of 1.2 nm. **c**, Typical Raman spectra of as-grown, ads-Ar and des-Ar 1L MoS$_2$. The characteristic peaks of as-grown, ads-Ar and des-Ar MoS$_2$ show no obviously different. **d**, PL spectra of as-grown, ads-Ar, partially des-Ar and des-Ar 1L MoS$_2$. Inset is the optical image of triangular 1L MoS$_2$ grains on sapphire. **e**, Typical SEM image of individual SWCNT on Si\SiO$_2$, the average distance between different SWCNT is <10 μm.